\documentclass[onecolumn,english,aps,prl,amsmath,amssymb,superscriptaddress]{revtex4}
\usepackage[latin9]{inputenc}
\usepackage{color}
\usepackage{amstext}
\usepackage{amssymb}
\usepackage{graphicx}
\usepackage[labelfont=bf]{caption}
\makeatletter
\@ifundefined{textcolor}{}
{%
 \definecolor{BLACK}{gray}{0}
 \definecolor{WHITE}{gray}{1}
 \definecolor{RED}{rgb}{1,0,0}
 \definecolor{GREEN}{rgb}{0,1,0}
 \definecolor{BLUE}{rgb}{0,0,1}
 \definecolor{CYAN}{cmyk}{1,0,0,0}
 \definecolor{MAGENTA}{cmyk}{0,1,0,0}
 \definecolor{YELLOW}{cmyk}{0,0,1,0}
}

\usepackage{color}

\newcommand{\beq}{\begin{eqnarray}}
\newcommand{\eeq}{\end{eqnarray}}

\makeatother
\usepackage{babel}

\begin{document}

\title{
Superconductivity modulated by structural phase transitions in pressurized vanadium-based kagome metals  
}

\author{Feng Du}

\thanks{These authors contributed equally to this work.}

\affiliation{Center for Correlated Matter and Department of Physics, Zhejiang University, Hangzhou 310058, China}
\affiliation  {Zhejiang Province Key Laboratory of Quantum Technology and Device, Department of Physics, Zhejiang University, Hangzhou 310058, China}

\author{Rui Li}

\thanks{These authors contributed equally to this work.}

\affiliation{Center for Correlated Matter and Department of Physics, Zhejiang University, Hangzhou 310058, China}
\affiliation  {Zhejiang Province Key Laboratory of Quantum Technology and Device, Department of Physics, Zhejiang University, Hangzhou 310058, China}

\author{Shuaishuai Luo}

\thanks{These authors contributed equally to this work.}

\affiliation{Center for Correlated Matter and Department of Physics, Zhejiang University, Hangzhou 310058, China}
\affiliation  {Zhejiang Province Key Laboratory of Quantum Technology and Device, Department of Physics, Zhejiang University, Hangzhou 310058, China}

\author{Yu Gong}

\affiliation{Beijing Synchrotron Radiation Facility, Institute of High Energy Physics, Chinese Academy of Sciences, Beijing 100049, China}

\author{Yanchun Li}

\affiliation{Beijing Synchrotron Radiation Facility, Institute of High Energy Physics, Chinese Academy of Sciences, Beijing 100049, China}

\author{Sheng Jiang}
\affiliation{Shanghai Institute of Applied Physics, Chinese Academy of Sciences, Shanghai 201204, China}

\author{Brenden R. Ortiz}

\affiliation{Materials Department and California Nanosystems Institute, University of California Santa Barbara, Santa Barbara, CA, 93106, United States}

\author{Yi Liu}

\affiliation{Key Laboratory of Quantum Precision Measurement of Zhejiang Province, Department of Applied Physics, Zhejiang University of Technology, Hangzhou 310023, China}

\author{Xiaofeng Xu}

\affiliation{Key Laboratory of Quantum Precision Measurement of Zhejiang Province, Department of Applied Physics, Zhejiang University of Technology, Hangzhou 310023, China}

\author{Stephen D. Wilson}

\affiliation{Materials Department and California Nanosystems Institute, University of California Santa Barbara, Santa Barbara, CA, 93106, United States}

\author{Chao Cao}

\affiliation{Center for Correlated Matter and Department of Physics, Zhejiang University, Hangzhou 310058, China}
\affiliation  {Zhejiang Province Key Laboratory of Quantum Technology and Device, Department of Physics, Zhejiang University, Hangzhou 310058, China}

\author{Yu Song}

\email{yusong$_$phys@zju.edu.cn}

\affiliation{Center for Correlated Matter and Department of Physics, Zhejiang University, Hangzhou 310058, China}
\affiliation  {Zhejiang Province Key Laboratory of Quantum Technology and Device, Department of Physics, Zhejiang University, Hangzhou 310058, China}

\author{Huiqiu Yuan}

\email{hqyuan@zju.edu.cn}

\selectlanguage{english}%

\affiliation{Center for Correlated Matter and Department of Physics, Zhejiang University, Hangzhou 310058, China}
\affiliation  {Zhejiang Province Key Laboratory of Quantum Technology and Device, Department of Physics, Zhejiang University, Hangzhou 310058, China}
\affiliation  {State Key Laboratory of Silicon Materials, Zhejiang University, Hangzhou 310058, China}

\begin{abstract}
The interplay of superconductivity with electronic and structural instabilities on the kagome lattice provides a fertile ground for the emergence of unusual phenomena.
The vanadium-based kagome metals $A$V$_3$Sb$_5$ ($A=$~K, Rb, Cs) exhibit superconductivity on an almost ideal kagome lattice, with the superconducting transition temperature $T_{\rm c}$ forming two domes upon pressure-tuning. The first dome arises from the competition between superconductivity and a charge-density-wave, whereas the origin for the second dome remains unclear. Herein, we show that the appearance of the second superconducting dome in KV$_3$Sb$_5$ and RbV$_3$Sb$_5$ is associated with transitions from hexagonal $P6$/$mmm$ to monoclinic $P2$/$m$ structures, evidenced by splitting of structural peaks from synchrotron powder X-ray diffraction experiments and imaginary phonon frequencies in first-principles calculations. In KV$_3$Sb$_5$, transition to an orthorhombic $Pmmm$ structure is further observed for pressure $p\gtrsim20$~GPa, and is correlated with the strong suppression of $T_{\rm c}$ in the second superconducting dome. Our findings indicate distortions of the crystal structure modulates superconductivity in $A$V$_3$Sb$_5$ under pressure, providing a platform to study the emergence of superconductivity in the presence of multiple structural instabilities.
\end{abstract}

\maketitle

\section*{Introduction}

The tuning of quantum materials provides a key for understanding their physics and a route towards potential applications. Highly tunable physical properties often emerge in the presence of electronic or structural instabilities, as seen in strongly correlated and two-dimensional materials \cite{Keimer2015,Si2010,Shibauchi2014,Cao2018}. Recently, kagome metals came into focus, because they natively harbor flat bands, saddle points and topological electronic structures \cite{Guo2009,Mazin2014,Ye2018,Liu2018,Kang2019}, fostering competing ground states and tunable physical properties. The kagome metals $A$V$_3$Sb$_5$ ($A=$~K, Rb, Cs) \cite{Ortiz2019,Ortiz2020,yin2021superconductivity,ortiz2020superconductivity} exhibit a giant anomalous Hall effect \cite{Yang2020,Yu2021_PRB} in the absence of local magnetic moments \cite{kenney2020absence}, which may result from a chiral charge-density-wave (CDW) that breaks time-reversal symmetry \cite{jiang2020discovery,Feng2021,mielke2021timereversal,yu2021evidence,setty2021electron}. Moreover, superconductivity with evidence for a fully-opened gap emerges at low-temperatures \cite{duan2021nodeless,mu2021swave,xu2021multiband}, and raises the possibility of topological superconductivity \cite{wang2020proximityinduced,liang2021threedimensional}. 

As a clean and powerful tool that can be utilized to tailor materials\textsc{\char13} atomic distances, and thus crystal and electronic structures, pressure is widely used to tune the properties of materials. The $A$V$_3$Sb$_5$ kagome metals are highly tunable by pressure or strain \cite{zhao2021nodal,Yu2021,chen2021double,Du2021,ChenCPL,Zhang2021,zhu2021doubledome,Du2021_Rb,yin2021strainsensitive,song2021competing,qian2021revealing}, 
and display complex evolution of superconductivity with pressure.
Under pressure, the CDW is quickly suppressed at $p_{\rm c}\lesssim 2$~GPa, forming a superconducting dome with $T_{\rm c}$ maximized at $p_{\rm c}$ \cite{Yu2021,chen2021double,Du2021,Du2021_Rb}, consistent with competing CDW and superconductivity. More strikingly, a second superconducting dome is observed at higher pressures in all variants of $A$V$_3$Sb$_5$ \cite{zhao2021nodal,Du2021,ChenCPL,Zhang2021,zhu2021doubledome,Du2021_Rb}, with proposed origins including a Liftshitz transition \cite{ChenCPL}, the formation of a three-dimensional structural network via Sb-Sb bonding \cite{tsirlin2021anisotropic}, an interplay with magnetism \cite{zhang2021firstprinciples}, and a pressure-induced phase with resistive anomalies \cite{Du2021,Du2021_Rb}. Similar two-dome superconductivity under pressure have been detected in unconventional superconductors, with their origins revealing integral aspects of the physics in these materials. For instance, in CeCu$_2$Si$_2$ the two domes result from distinct pairing due to spin or valence fluctuations \cite{Yuan2003,Yuan2006}, and in iron chalcogenides they are tied to a change from Fermi-liquid to non-Fermi-liquid \cite{Sun2012,Sun2018}. Therefore, elucidating the underlying mechanism for the two superconducting domes under pressure
in the $A$V$_3$Sb$_5$ series may facilitate further understanding and potential functional engineering of these highly tunable kagome superconductors.



Here, combining synchrotron powder X-ray diffraction (XRD) and {\it ab initio} calculations, we show that transitions from hexagonal $P6$/$mmm$ to monoclinic $P2$/$m$ structures occur in both KV$_3$Sb$_5$ and RbV$_3$Sb$_5$. By comparing with resistivity measurements carried out using the same pressure medium, these structural transitions are identified to coincide with appearance of the second superconducting dome in the two compounds.
In KV$_3$Sb$_5$, another transition into an orthorhombic $Pmmm$ structure is observed upon further increase of pressure, whose emergence is associated with a strong suppression of $T_{\rm c}$. 
These findings highlight the role of pressure-induced structural modulations in the complex evolution of superconductivity in $A$V$_3$Sb$_5$ under pressure, and underscore the effects of structural changes on properties of layered materials under high pressure.

\section*{Results}

\begin{figure*}
	\includegraphics[width=0.9\columnwidth]{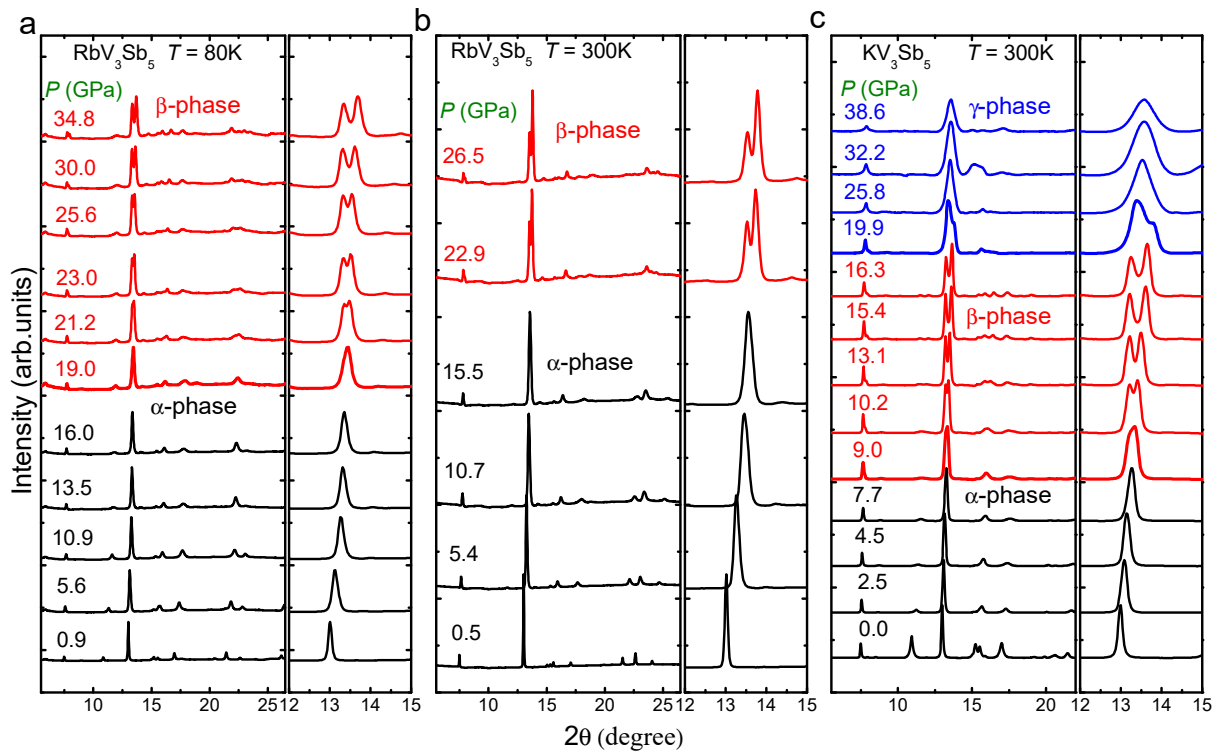} \protect\caption{{\bf Powder X-ray diffraction patterns under pressure.} XRD data under various pressures for (a) RbV$_3$Sb$_5$ at 80~K, (b) RbV$_3$Sb$_5$ at 300~K, and (c) KV$_3$Sb$_5$ at 300~K. The XRD patterns have been shifted vertically for clarity. The right side of each panel shows zoomed-in data, focusing on the peak near 13$^\circ$.}
	\label{Fig_XRD_raw}
\end{figure*}

\subsection*{Powder X-ray diffraction under pressure}

At ambient pressure, $A$V$_3$Sb$_5$ forms a hexagonal $P6$/$mmm$ structure above the CDW transition temperature, with a V kagome lattice interwoven with a Sb hexagonal lattice in the same layer, further encapsulated by layers of Sb (that form honeycomb lattices) above and below. These V$_3$Sb$_5$ slabs are well-separated by alkaline metals ions, resulting in a highly two-dimensional structure. To investigate how the crystal structure evolves under pressure and its relation to superconductivity, we carried out systematic powder XRD measurements under pressure. To facilitate a direct comparison with phase diagrams determined using resistivity \cite{Du2021,Du2021_Rb}, the same liquid pressure medium (silicon oil) is used in the XRD measurements.

XRD data for RbV$_3$Sb$_5$ and KV$_3$Sb$_5$ under pressure are shown in Fig.~\ref{Fig_XRD_raw} [see Supplementary Fig. 1(a) for additional data on KV$_3$Sb$_5$].
Diffraction patterns consistent with the ambient pressure hexagonal structure ($\alpha$-phase) are found to persist in RbV$_3$Sb$_5$ up to $\approx16$~GPa [Figs.~\ref{Fig_XRD_raw}(a), (b)], and in KV$_3$Sb$_5$ up to $\approx8$~GPa [Fig.~\ref{Fig_XRD_raw}(c)]. Above these pressures, qualitative changes occur in the diffraction patterns, most prominently the peak at $\approx13^\circ$ broadens and splits into two, with the split peak at larger scattering angles being more intense.
The observation of such a splitting provides unequivocal evidence for a pressure-induced structural phase transition into a new phase ($\beta$-phase). It is interesting to note that the splitting may become difficult to observe, if the sample has prominent preferred orientations [see Supplementary Fig.~1(b) for details]. The broadening first appears around 19.0~GPa in RbV$_3$Sb$_5$ [Fig.~\ref{Fig_XRD_raw}(a)] and around 9.0~GPa in KV$_3$Sb$_5$ [Fig.~\ref{Fig_XRD_raw}(c)], with the larger pressure in RbV$_3$Sb$_5$ consistent with Rb atoms contribution to a negative chemical pressure. 
In addition, the evolution of the diffraction pattern for RbV$_3$Sb$_5$ at 300~K is consistent with that of 80~K, indicating the pressure for the $\alpha$-$\beta$ phase transition does not vary strongly between the two temperatures. 

\begin{figure*}
	\includegraphics[width=0.9\columnwidth]{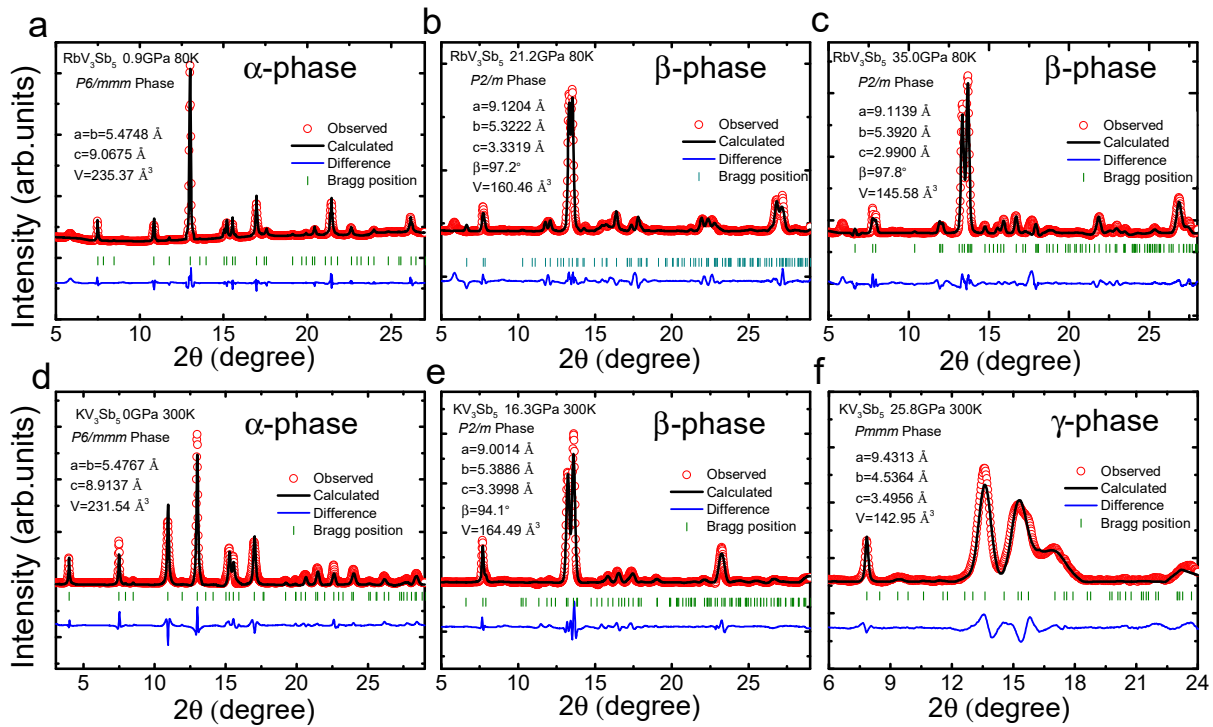} \protect\caption{{\bf Fitting of representative diffraction patterns.} XRD data for (a)-(c) RbV$_3$Sb$_5$ and (d)-(f) KV$_3$Sb$_5$ under various pressures. The diffraction data (open red circles) are fit to a hexagonal $P6$/$mmm$, monoclinic $P2$/$m$ or orthorhombic $Pmmm$ structure, using the Le Bail method.}
	\label{Fig_XRD_fit}
\end{figure*}

While the $\beta$-phase persists to the highest measured pressure in RbV$_3$Sb$_5$ (34.8~GPa), the split peaks merge into a single peak in KV$_3$Sb$_5$ for $p\gtrsim20$~GPa [Fig.~\ref{Fig_XRD_raw}(c) and Supplementary Fig.~1(a)], signifying the appearance of a new structural phase ($\gamma$-phase). 
At 19.9~GPa [Fig.~\ref{Fig_XRD_raw}(c)], even though two peaks are still present near 13$^\circ$, the peak at smaller angles becomes more intense, distinct from scattering patterns of the $\beta$-phase. On the other hand, the more intense peak of the two closely track the single peak in the $\gamma$-phase. Therefore, the two peaks at 19.9~GPa can be understood to result from the sample being in a mixture of $\beta$- and $\gamma$-phases, which suggests the transition to be first-order-like and occurs close to 19.9~GPa.

To understand the nature of structural phases under pressure in RbV$_3$Sb$_5$ and KV$_3$Sb$_5$, the diffraction patterns are analyzed using the Le Bail method, with representative fits shown in Fig.~\ref{Fig_XRD_fit}. Peaks in the $\alpha$-phase can be indexed by the ambient pressure hexagonal $P6$/$mmm$ structure [Figs.~\ref{Fig_XRD_fit}(a) and (d)], with the intense peak around 13$^\circ$ corresponding to the in-plane Bragg peak (110).
Diffraction patterns of the $\beta$-phase can be described by a monoclinic $P2$/$m$ structure [Figs.~\ref{Fig_XRD_fit}(b),(c) and (e)], which fully accounts for the split peaks around 13$^\circ$, and are indexed as (020) and (011) of the monoclinic unit cell. 
For the $\gamma$-phase of KV$_3$Sb$_5$, the diffraction patterns can be described by an orthorhombic $Pmmm$ structure [Fig.~\ref{Fig_XRD_fit}(f)], with a single intense (111) peak (indexed in the orthorhombic unit cell) around 13$^\circ$. 
Lattice parameters for the $\alpha$-, $\beta$- and $\gamma$-phases under representative pressures are shown in Fig.~\ref{Fig_XRD_fit} (see Supplementary Figs.~2-4 for detailed evolution of the lattice parameters). 

\begin{figure}
	\includegraphics[width=0.7\columnwidth]{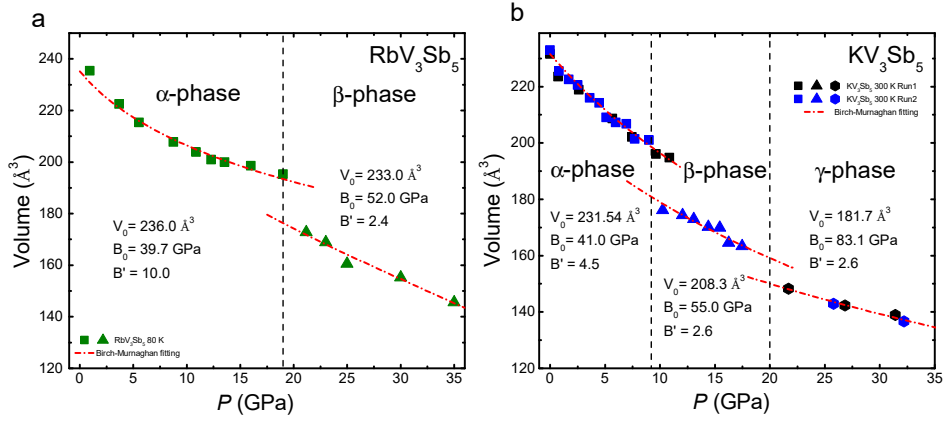} \protect\caption{{\bf Evolution of the unit cell volume as a function of pressure.} The volume of the unit cell as a function of applied pressure for (a) RbV$_3$Sb$_5$, and (b) KV$_3$Sb$_5$. The dashed lines are fits to the Birch-Murnaghan equation of state.}
	\label{Fig_volume}
\end{figure}

Evolution of the unit cell volume with pressure is summarized in Fig.~\ref{Fig_volume}. Within each structural phase, pressure-dependence of the unit cell volume is well-described using the 3rd-order Birch-Murnaghan equation of states (dashed lines in Fig.~\ref{Fig_volume}). On the other hand, a single set of parameters cannot describe the evolution of the unit cell volume over the entire measured pressure range, further corroborating the presence of distinct structural phases under pressure. 
A large $\sim9$\% volume reduction is observed across the $\alpha$-$\beta$ phase transition in both RbV$_3$Sb$_5$ and KV$_3$Sb$_5$, 
and a further $\sim6$\% volume reduction is seen across the $\beta$-$\gamma$ phase transition in KV$_3$Sb$_5$. 

\subsection*{First-principles calculations of phonon spectra}

\begin{figure}
	\includegraphics[width=0.9\columnwidth]{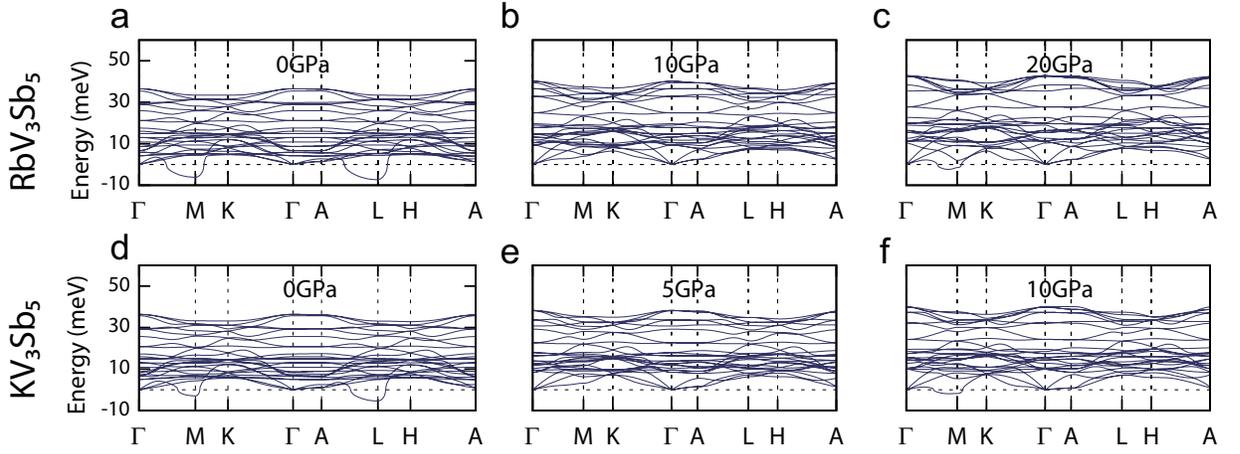} \protect\caption{{\bf Structural stability studied via first-principles calculations.} (a)-(c) Calculated phonon spectra for RbV$_3$Sb$_5$ under various pressures. (d)-(f) Calculated phonon spectra for KV$_3$Sb$_5$ under various pressures.}
	\label{Fig_phonon}
\end{figure}

The phonon spectra of RbV$_3$Sb$_5$ and KV$_3$Sb$_5$ are calculated from first-principles using a hexagonal $P6$/$mmm$ unit cell, with results summarized in Fig.~\ref{Fig_phonon}. The negative phonon frequencies in the plots indicate imaginary phonon modes, corresponding to structural instabilities. The reciprocal lattice vector and the symmetry classification of the lowest phonon mode can provide information on the ground state crystal structure.
At zero pressure, imaginary phonon modes are found at the M and L points in both RbV$_3$Sb$_5$ [Fig.~\ref{Fig_phonon}(a)] and KV$_3$Sb$_5$ [Fig.~\ref{Fig_phonon}(d)], with the leading instability at the L point. These phonon instabilities are B$_{1u}$ modes, corresponding to a CDW state with a star of David (or its inverse) deformation of the kagome lattice \cite{ortiz2021fermi}, consistent with previous calculations for $A$V$_3$Sb$_5$ \cite{zhang2021firstprinciples,Tan2021}. Increasing pressure to 10~GPa in RbV$_3$Sb$_5$ and 5~GPa in KV$_3$Sb$_5$ [Figs.~\ref{Fig_phonon}(b) and (e)], the imaginary phonons associated with the CDW disappear, which indicates the hexagonal $P6$/$mmm$ structure is stable at these pressures, consistent with the experimentally observed suppression of CDW above $p_{\rm c}$ ($\lesssim2$~GPa) \cite{Yu2021,chen2021double,Du2021,Du2021_Rb}. 

With further increase of pressure, new imaginary phonon modes emerge around M point in both RbV$_3$Sb$_5$ and KV$_3$Sb$_5$ [Figs.~\ref{Fig_phonon}(c) and (f)], indicating the appearance of another structural instability. These instabilities are B$_{3g}$ modes, and are absent around the L point. Therefore, this structural instability is distinct from the CDW that occurs below $p_{\rm c}$, and can be instead associated with the $\beta$-phase uncovered in our XRD measurements. It is noteworthy that while 10~GPa is sufficient to induce this structural instability in KV$_3$Sb$_5$ [Fig.~\ref{Fig_phonon}(f)], no imaginary phonon is found in RbV$_3$Sb$_5$ at the same pressure [Fig.~\ref{Fig_phonon}(b)], consistent with the experimental observation that the $\beta$-phase appears at a higher pressure in RbV$_3$Sb$_5$. 

A previous first-principles study also found a second structural instability for CsV$_3$Sb$_5$ around 30-35~GPa, which is associated with a B$_{1u}$ phonon mode, similar to the ambient pressure CDW \cite{zhang2021firstprinciples}. The structural instability found in our calculations for RbV$_3$Sb$_5$ and KV$_3$Sb$_5$ is different, because (1) it is associated with a B$_{3g}$ phonon mode, and (2) the instability is not limited to a small pressure range, but persists to the highest pressures in our calculations (40~GPa for RbV$_3$Sb$_5$ and 30~GPa for KV$_3$Sb$_5$, see Supplementary Fig.~5 for calculated phonon spectra at higher pressures). Our first-principles calculations therefore point to a robust structural instability in pressurized RbV$_3$Sb$_5$ and KV$_3$Sb$_5$, in contrast to results from similar calculations for CsV$_3$Sb$_5$ under pressure \cite{zhang2021firstprinciples}, but consistent with our experimental XRD results.

\section*{Discussion and Conclusion}
\begin{figure}
	\includegraphics[width=0.8\columnwidth]{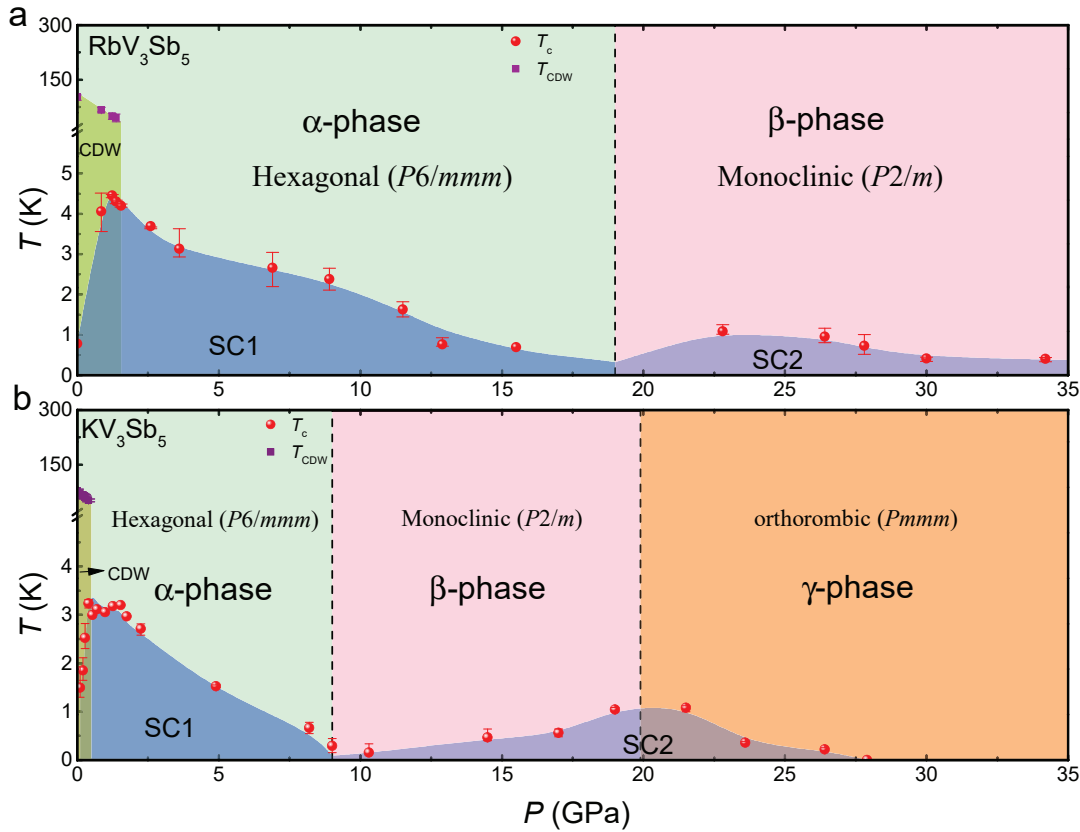} \protect\caption{{\bf Pressure-temperature phase diagrams.} (a) $P-T$ phase diagram for RbV$_3$Sb$_5$. (b) $P-T$ phase diagram for KV$_3$Sb$_5$. SC1 and SC2 correspond to the two superconducting domes revealed in transport measurements \cite{Du2021_Rb,Du2021}. $T_{\rm c}$ and the CDW ordering temperature $T_{\rm CDW}$ for RbV$_3$Sb$_5$ are from Ref.~\cite{Du2021_Rb}, and those for KV$_3$Sb$_5$ are from Ref.~\cite{Du2021}. The vertical dashed lines separate distinct structural phases, and are determined from the lowest pressure that the new phases are observed.}
	\label{Fig_phase_diagram}
\end{figure}

Based on the XRD results in this work and previous resistivity measurements \cite{Du2021,Du2021_Rb}, the temperature-pressure phase diagrams for RbV$_3$Sb$_5$ and KV$_3$Sb$_5$ are constructed and shown in Fig.~\ref{Fig_phase_diagram}. As can be seen, the most striking feature in both RbV$_3$Sb$_5$ and KV$_3$Sb$_5$ is the observation of a structural transition from the $\alpha$-phase to the $\beta$-phase coinciding with a minima in $T_{\rm c}(p)$, which separates the two superconducting domes. Therefore, while the first superconducting dome arises from a competition between superconductivity and the CDW, the present XRD measurements and first-principles calculations provide a convincing case for the transition between two structural phases as the cause for two-dome superconductivity in these materials. The evolution of phonons and their coupling to electrons across different structural phases likely play important roles in the evolution of superconductivity. 

Appearance of the $\gamma$-phase in KV$_3$Sb$_5$ is accompanied by a strong reduction in $T_{\rm c}$, which becomes fully suppressed above $\approx30$~GPa \cite{Du2021}. In RbV$_3$Sb$_5$, superconductivity becomes gradually suppressed with increasing pressure above $\approx25$~GPa. Nonetheless, the suppression of superconductivity is much less pronounced compared to KV$_3$Sb$_5$ \cite{Du2021_Rb}, and no $\gamma$-phase is detected up to $\approx35$~GPa. These behaviors suggest that the $\gamma$- and $\beta$-phases are close in energy, with the former less favorable for superconductivity. 
Alternatively, the suppression of $T_{\rm c}$ in the second superconducting dome may be related to clear anomalies in the normal state resistivity, found in both RbV$_3$Sb$_5$ \cite{Du2021_Rb} and KV$_3$Sb$_5$ \cite{Du2021}. 

Due to difficulties in accurately obtaining diffraction peak intensities under pressure, Rietveld analysis cannot be carried out on our XRD data to solve crystals structures of the $\beta$- and $\gamma$-phases. Nonetheless, insights on these structures can be gleaned by examining the structural evolution of related layered materials under pressure. Cold-pressed graphite is found to transition from a layered hexgonal structure to a monoclinic $C2$/$m$ or an orthorhombic $Pnma$ structure under pressure \cite{Li2009,Wang2011,Wang2012}, with the formation of covalent bonds between distorted graphite layers. 
For 1T-TaS$_2$ under pressure, the layered hexagonal $P6_3$/$mmc$ structure first distorts, then interlayer S-S bonds form, and finally evolves into a tetragonal $I4$/$mmm$ structure \cite{Dong2021}, possibly via a monoclinic $C2$/$m$ intermediate phase \cite{Wang2020_TaS2}. A common theme in these materials is that the structures under ambient pressure are highly two-dimensional, and with the application of pressure interlayer bonds gradually form, before finally transitioning into a three-dimensional structure. These three-dimensional structures stabilized under pressure are often monoclinic, although orthorhomic or tetragonal structures may be close in energy. These considerations suggest that the $\beta$-phase is likely a three-dimensional structure, with distorted kagome planes and interlayer Sb-Sb bonds, while the $\gamma$-phase is close in energy, reminiscent of monoclinic and orthorhombic carbon \cite{Li2009,Wang2011,Wang2012}. This proposal is consistent with the large reduction in the unit cell volume across the $\alpha$-$\beta$ phase transition, which is likely dominated by reduced interlayer atomic distances.


In the case of CsV$_3$Sb$_5$, resistivity measurements suggest the presence of two superconducting domes, similar to RbV$_3$Sb$_5$ and KV$_3$Sb$_5$. However, powder XRD measurements at room temperature indicate a hexagonal $P6$/$mmm$ structure ($\alpha$-phase) is adopted up to $\approx40$~GPa \cite{zhang2021pressureinduced}, seemingly different from our findings in RbV$_3$Sb$_5$ and KV$_3$Sb$_5$. Nonetheless, for pressures associated with the second superconducting dome, Raman scattering detected sharp changes in the zone-center optical phonons \cite{chen2021double} and single crystal XRD revealed the formation of interlayer Sb-Sb bonds \cite{tsirlin2021anisotropic}, both already present at room temperature. Therefore, although CsV$_3$Sb$_5$ maintains a hexagonal $P6$/$mmm$ structure at room temperature up to 40~GPa, significant structural modulations occur under pressures associated with the second superconducting dome. 
Moreover, whether these structural distortions would lead to a $\beta$-phase upon cooling below room temperature in CsV$_3$Sb$_5$, requires further low-temperature XRD measurements under pressure.

In conclusion, we systematically investigated the evolution of RbV$_3$Sb$_5$ and KV$_3$Sb$_5$ under pressure using powder X-ray diffraction and first-principles calculations, and uncovered a transition into a distinct structural phase upon increasing pressure. Compared against resistivity measurements obtained using the same pressure medium, we find the structural transition coincides with a minima in $T_{\rm c}$, and naturally accounts for the presence of two superconducting domes in these materials. In KV$_3$Sb$_5$, an additional structural phase transition is found at higher pressures, which is associated with a strong suppression of superconductivity in the second superconducting dome. These findings demonstrate an integral role for structural instabilities in the complex evolution of superconductivity under pressure in the $A$V$_3$Sb$_5$ vanadium kagome metals, and suggest a potential role of the corresponding soft phonons in determining their physical properties under ambient pressure. 

\section*{Methods}


\subsection*{Powder X-ray diffraction under pressure}
Single crystals of RbV$_3$Sb$_5$ and KV$_3$Sb$_5$ were grown using a self-flux method, with detailed procedures documented in Refs.~\cite{Ortiz2019,Ortiz2020}. Powder XRD measurements were performed using the 4W2 beamline at the Beijing Synchrotron Radiation Facility (BSRF) and the BL15U1 beamline at the Shanghai Synchrotron Radiation Facility (SSRF), both using a monochromatic beam of wavelength 0.6199~\AA. The pressed powder samples were cut into pieces with dimensions of approximately $50\times50~\times30\mu$m$^3$, and loaded into diamond anvil cells (DACs) with culet sizes of $300~\mu$m in diameter. Stainless steel T301 gaskets were preindented to about $40~\mu$m thick with a hole of $140~\mu$m in diameter drilled as the sample chamber. Silicon oil was used as the pressure-transmitting medium. The DACs were loaded together with several small ruby balls for pressure determination using the ruby fluorescence method. 

Samples used in XRD measurements were obtained by crushing single crystals. Two XRD runs on RbV$_3$Sb$_5$ were carried out, one at 80~K [Fig.~\ref{Fig_XRD_raw}(a)] and one at 300~K [Fig.~\ref{Fig_XRD_raw}(b)]. The pressure pathways for these two runs are from 0.9~GPa to 34.8~GPa and from 0.5~GPa to 26.5~GPa, respectively. For the run at 80~K, crushed single crystals inside the DAC were cooled in a liquid nitrogen cryostat, and powder XRD patterns were collected through double Mylar windows of the cryostat. Two XRD runs on KV$_3$Sb$_5$ were carried out. The pressure pathways for these two runs are from ambient pressure to 38.6~GPa [run 1 shown in Fig.~\ref{Fig_XRD_raw}(c)] and to 31.4~GPa [run 2 shown in Supplementary Fig.~1(a)], respectively. The \texttt{Fit2d} program was used for image integration and the XRD patterns were fit using the \texttt{Fullprof} program with the Le Bail method.

Evolution of the unit cell volume with pressure was fit to the third-order Birch-Murnaghan equation of states \cite{PhysRev.71.809}:
\begin{equation*}\label{key}
P = \frac{3}{2}B_0[(\frac{V_0}{V})^{\frac{7}{3}}-(\frac{V_0}{V})^{\frac{5}{3}}] \times \left\{1+\frac{3}{4}({B_0^\prime}-4)[(\frac{V_0}{V})^{\frac{2}{3}}-1]\right\},
\end{equation*}
where $V_0$ is the unit cell volume under ambient pressure. $B_0$ and $B_0^\prime$ respectively represent the bulk modulus and first-order derivative of the bulk modulus at ambient pressure.

\subsection*{First-principles phonon calculations}
The first-principles calculations of phonons were carried out based on density functional theory and density functional perturbation theory, as implemented in \texttt{QUANTUM ESPRESSO} \cite{Giannozzi2009}. The Perdew, Burke and Ernzerho parameterization of the general gradient approximation was used for the exchange-correlation functionals \cite{Perdew1996}. The energy cutoff was chosen to be 88 Ry, and the Brillouin zone was sampled with a $12\times12\times6$ $k$-mesh to ensure the convergence of our calculations. The phonon dispersion and electron-phonon coupling were calculated with a $3\times3\times2$ $q$-mesh. Both the lattice parameters and the atomic coordinates were optimized so that the forces on each atom were smaller than 0.0001 Ry/{\AA} and internal stress less than 0.1~kbar. In addition, the phonon dispersions were verified by additional calculations using the finite displacement method, as implemented in the \texttt{phonopy} code \cite{Togo2015}.

\section*{Data availability}
All data supporting the findings of this work are available from the corresponding authors upon reasonable request.

\section*{Acknowledgments}

We acknowledge Zhongyi Lu for helpful discussions and Guanghan Cao for assistance in sample preparation. This work was supported by the National Key R\&D Program of China (No. 2017YFA0303100, No. 2016YFA0300202, and No. 2017YFA0403401), the Key R\&D Program of Zhejiang Province, China (2021C01002), the National Natural Science Foundation of China (No. 12034017, No. 11974306, No. 11874137 and No. 11974061), and the Fundamental Research Funds for the Central Universities of China.  S.D.W. and B.R.O. gratefully acknowledge support via the UC Santa Barbara NSF Quantum Foundry funded via the Q-AMASE-i program under award DMR-1906325.  B.R.O. also acknowledges support from the California NanoSystems Institute through the Elings fellowship program.


\bibliography{bibfile}
\end{document}